\documentclass{jpsj2}
\title{Enhancement of magnetic order in the incommensurate phase of 
Mg-doped CuGeO$_{3}$}

\author{C.\textsc{Stock}$^{1}$, S. \textsc{Wakimoto}$^{1}$, R.J. \textsc{Birgeneau}$^{1}$, S. \textsc{Danilkin}$^{2}$, J. \textsc{Klenke}$^{2}$, P. \textsc{Smeibidl}$^{2}$, and P. \textsc{Vorderwisch}$^{2}$}

\inst{$^{1}$ Department of Physics, University of Toronto, Ontario, Canada 
M5S 1A7 \\
$^{2}$ BENSC, Hahn-Meitner Institute, D-14109, Berlin, Germany}

\abst{We present elastic neutron scattering results at high magnetic fields for Mg-doped CuGeO$_{3}$.  For low magnesium concentrations, where a spin-Peierls phase is present, we find an enhancement of the long range antiferromagnetic order in the structural incommensurate phase induced by a high magnetic field.  In the incommensurate phase, the N\'eel temperature increases with field leading to a gain in the magnetic Bragg peak intensity at low temperatures.  We also find that the magnetic Bragg peaks remain commensurate even though the structure is incommensurate at high magnetic fields due to the formation of solitons.  We interpret these results in terms of the solitons formed at high-magnetic fields effectively acting as doped impurities.   Our results point to a strongly disordered ground state with the the antiferromagnetic order localized around impurity sites.}

\kword{spin-Peierls, neutron scattering, doping, CuGeO$_{3}$}

\begin{document}
\maketitle

\section{Introduction} 

	Impurity doping of low-dimensional magnetic systems often results in non-intuitive ground states and critical behavior differing from that of the pure or parent material.  In this respect, doped CuGeO$_{3}$, with Cu$^{2+}$ ions forming a one-dimensional (1D) chain network along the $c$-axis, has generated much interest recently. CuGeO$_{3}$ was the first inorganic material discovered to undergo a spin-Peierls transition.~\cite{Hase93:70}  In the Peierls state the copper atoms form dimer pairs and a gap opens in the magnetic excitation spectrum.  The spin coupling is dominantly along the chain direction [001] with $J_{c} = 10.5$~meV, compared to couplings of -0.1 meV and 1.0 meV perpendicular to the chains along the $a$ and $b$ directions, respectively.~\cite{Nishi94:50}  The excitations have been shown to be well modelled by a singlet ground state with an excited triplet state.~\cite{Cowley96:8}  Due to the strong 1D nature of the spin interactions, it is not expected that the spins should order at low temperatures, consistent with experimental results in the pure system.~\cite{Regnault96:53}

	One of the most unique aspects of CuGeO$_{3}$ is the ability to introduce impurities on different lattice sites in a very controlled manner.  This is a particular advantage of this inorganic system over previous organic systems where it was extremely difficult to dope impurities systematically.  The introduction of a small amount of impurities on either the Cu or the Ge site has been shown to result in a long-range antiferromagnetic (AF) ordered state at low temperatures.~\cite{Lussier95:7}  Regardless of the impurity and the dopant site the overall phase diagrams are qualitatively identical as a function of doping.  At low impurity dopings long-range AF order is observed to coexist with the dimerized spin-Peierls state until a critical concentration where the dimerized state is replaced with a uniform long-range magnetic ordered phase. Despite the apparent similarity between the phase diagrams of all doped impurities it has been found that non-magnetic impurities and spin-1 impurities produce slightly different ground states with the moment pointing along the [001] direction for the nonmagnetic case and the [100] direction in the case of a spin-1 impurity.~\cite{Coad96:8}

	The phase diagram at zero magnetic field for Cu$_{1-x}$Mg$_{x}$GeO$_{3}$ has been studied in detail with both x-ray and neutron scattering techniques.  In this system the doped nonmagnetic Mg$^{2+}$ ions occupy the Cu$^{2+}$ sites.  The critical properties of both the magnetic and structural distortions have been investigated by Nakao \textit{et al.} and Nishi \textit{et al.} using susceptibility and neutron diffraction measurements.~\cite{Nakao99:11,Nishi00:69}  These studies found the presence of a dimerized long-range AF ordered state at lower Mg concentrations and a uniform AF phase at higher concentrations with the absence of any spin-Peierls order or dimer structure.  The boundary of these phases takes place around $x_{c}$ $\sim$ 2.5~\% Magnesium doping.   The structural phase diagram has also been carefully studied using x-rays for a variety of impurities and concentrations.~\cite{Lumsden98:58,Wang99:83,Wang03:72,Kiryukhin00:61}  The excitation spectrum for the low temperature N\'eel state has also been mapped out and found to be in good agreement with theory for disordered spin-Peierls systems.~\cite{Gehring00:69}

	At high magnetic fields the dimerized phase becomes unstable to the formation of domain walls or solitons.    In the pure undoped material this is a first order transition which occurs at an applied magnetic field of 12.5 T and is characterized by the splitting of the spin-Peierls superlattice peaks into pairs of incommensurate peaks.~\cite{Kiryukhin95:52,Ronnow00:84} With doping this first order commensurate to incommensurate transition becomes a continuous or second order transition and has been studied in detail for Mg, Zn, and Ni doping using high resolution synchrotron x-ray scattering techniques.~\cite{Christianson02:66,Kiryukhin96:76}  Excitations above the incommensurate transition have been studied in the pure system and, in contrast to the soft modes in classical incommensurate structures, three gapped modes were clearly observed at high magnetic fields.~\cite{Enderle01:87}

	In this work, we study the effect of impurities on the high-magnetic field incommensurate phase using single crystals of Cu$_{1-x}$Mg$_{x}$GeO$_{3}$ with $x=0.02$ and $0.04$.  We find that the solitons formed at high magnetic fields essentially act as impurities breaking the chains and enhancing the magnetic order.  To confirm this picture we have compared our results at low Mg dopings ($x=0.02$) in the dimerized phase to those of higher dopings ($x=0.04$) where a spin-Peierls state is absent and a uniform antiferromagnet is formed. These results point to the impurity induced AF order being localized around impurity sites.

\begin{figure}[t]
\begin{center}
\includegraphics[width=8cm] {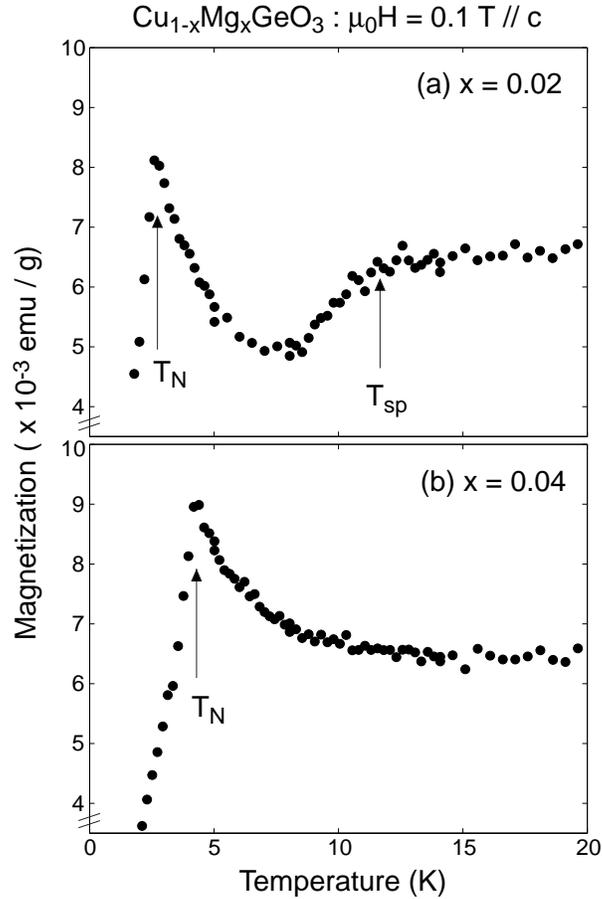}
\end{center}
\caption{\label{Neel} DC magnetization plotted as a function of 
temperature for both (a) $x=0.02$ and (b) $x=0.04$ samples.  The applied field is 0.1 T aligned along the \textit{c} direction.  The N\'eel and spin-Peierls transitions are labelled for the 2~\% sample.  The 4~\% only displays N\'eel order.}
\end{figure}

\section{Experimental Details}

	Single crystals of Cu$_{1-x}$Mg$_x$GeO$_{3}$ with $x=0.02$ and $0.04$ 
were grown using the floating zone method.~\cite{Revocolevschi99:198}  Stoichiometric concentrations of CuO, MgO, and GeO$_{2}$ were ground into powder, and then feed rods were prepared by annealing under O$_{2}$ flow at 900$^{\circ}$C for 12 hours.  The crystals were grown in an image furnace under O$_{2}$ flow along the $a$ direction with a growth rate of $\sim$ 0.5~mm/hour.  The grown crystals had a self-cleaving surface perpendicular to the growth direction. The crystals were 0.5~cc in volume and were characterized by measurements of the DC magnetization and heat capacity.  Details and discussion of the sample characterization are given in the next section.

	Neutron scattering experiments were conducted at the E1 thermal and V2 cold neutron spectrometers located at the BER II research reactor, HMI Berlin.  An Oxford vertical cryomagnet was used to reach temperatures as low as $\sim 1.6$~K and applied magnetic fields as high as $\mu_{0}H \sim 14$~T. The sample was mounted in the magnet with the cleaving face pointing down so that $\bf{Q}$ positions of the form $(0KL)$ lay within the scattering plane.  For both spectrometers a pyrolytic graphite (002) reflection was used as a monochromator and analyzer.  At the cold neutron instrument (V1) the incident neutron energy was set at $E_{i}=3.5$~meV and collimation fixed at 60'-S-60'-60' (S denotes sample).   The thermal spectrometer (E1) was configured with $E_{i}=13.9$~meV and a collimation of 40'-40'-S-40'-open.  To filter out higher order neutrons on the cold instrument a nitrogen cooled beryllium filter was used on the scattered side and on the thermal instrument a graphite filter was used on the incident side.  The lattice constants obtained at base temperature were $b=8.41$~\AA~ and $c=2.95$~\AA, corresponding to the reciprocal lattice units $b^{*}=0.75$~\AA$^{-1}$ and $c^{*}=2.13$~\AA$^{-1}$. 

\section{Sample Characterization}

\begin{figure}[t]
\begin{center}
\includegraphics[width=8cm] {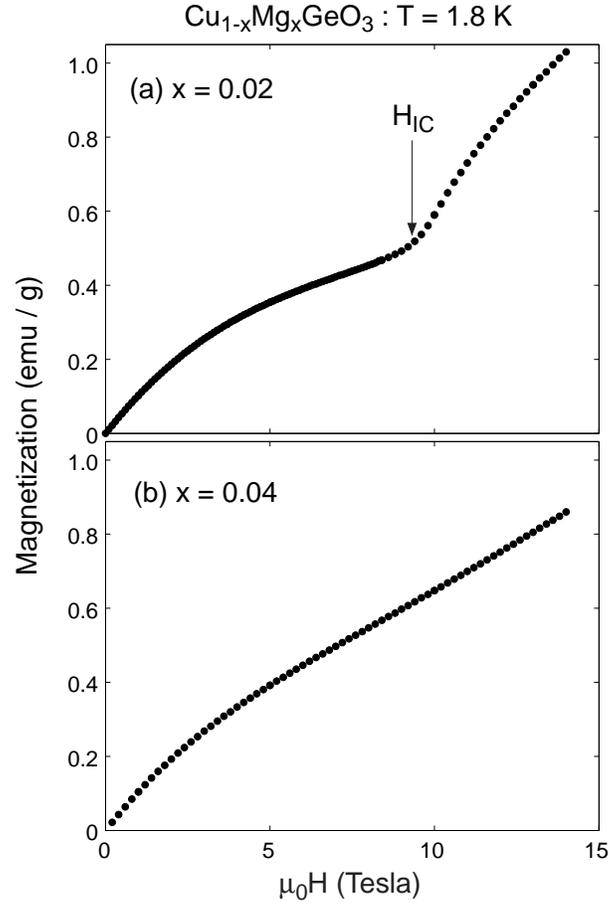}
\end{center}
\caption{\label{mag_H} DC magnetization plotted as a function of applied magnetic field up to 14 T.  The magnetic field is aligned along the \textit{a} direction.  The anomaly associated with the transition to the incommensurate phase is clearly visible in the 2~\% sample at $\sim$ 9.5 T, while no anomaly is observed in the 4~\% sample.}
\end{figure}

	The crystals were characterized based on the transition temperatures measured by magnetization and heat capacity.  DC Magnetization measurements are shown in Fig.~\ref{Neel} for both samples and were made using the Quantum Design ACMS system.  An applied magnetic field of $\mu_{0}H=0.1$~T was aligned along the [001] direction.  The $x=0.02$ sample shows both a broadened but clear spin-Peierls transition at around 11~K and a N\'eel temperature of 2.5~K indicating the coexistence between the two phases and presence of dimerized AF order.  The $x=0.04$ sample shows no sign of a spin-Peierls transition as only a clear N\'eel transition at 4.3 K is observed.  The physical properties of both of these samples are consistent with the previously established phase diagram.~\cite{Masuda00:284} To check the homogeneity of the samples, we have confirmed that the magnetization at both the top and bottom of the crystal rod, obtained from the floating zone furnace, were the same, indicating a very uniform Mg concentration throughout the crystal.  The data for these low field magnetization results are also consistent with previous studies using highly sensitive SQUID magnetometers.  

\begin{figure}[t]
\begin{center}
\includegraphics[width=8cm] {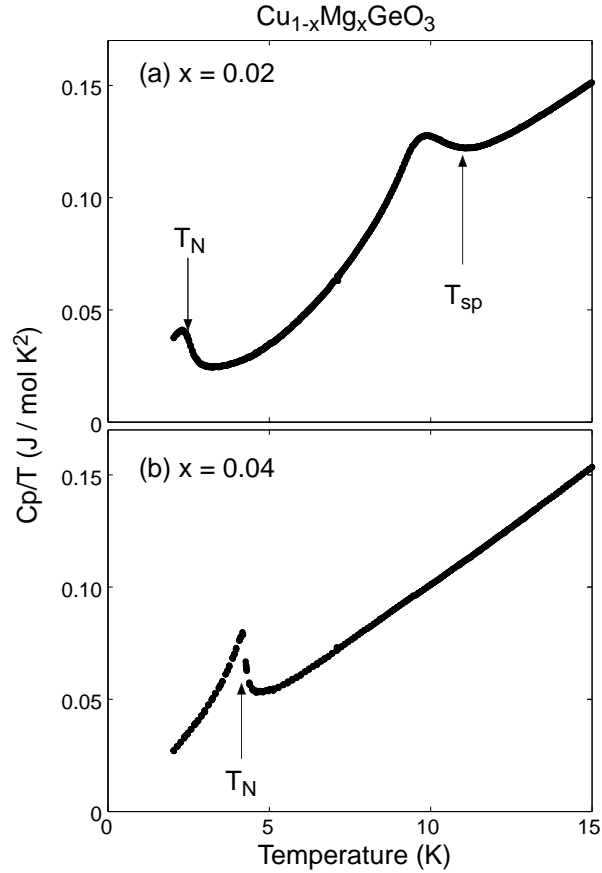}
\end{center}
\caption{\label{hc} Cp/T plotted as a function of temperature for both 
2~\% and 4~\% samples at zero applied magnetic field.  The 2~\% sample 
clearly shows both N\'eel and spin-Peierls transitions.  The 4~\% only 
shows a transition to N\'eel order.  The high temperature dependence is 
discussed in the text.}
\end{figure}

	From the magnetization as a function of applied field we clearly identify the commensurate-incommensurate phase transition.  Figure~\ref{mag_H} shows the DC magnetization as a function of applied field at 1.8 K, below the N\'eel temperature of the 2~\% sample.  For these measurements, the magnetic field was aligned along the [100] direction, the same field configuration as used in our neutron experiments.  The top panel of Fig.~\ref{mag_H} (2~\% sample) shows a clear break at 9.5~T.  This transition has been shown to be associated with the commensurate to incommensurate transition in previous studies in Si doped CuGeO$_{3}$ which shows a dimerized AF phase.~\cite{Grenier98:44}  The bottom panel of Fig.~\ref{mag_H} shows the DC magnetization as a function of applied field for the 4~\% sample with no sign of any break or transition at high magnetic fields.  This behavior is consistent with a uniform AF phase where the spin-Peierls phase, and hence the commensurate-incommensurate transition, is absent.

	Heat capacity at constant pressure, $C_{p}$, was measured using the Quantum Design heat capacity option, which utilizes the relaxation technique.  The results are displayed in Fig.~\ref{hc} with the 2~\% sample in the upper panel and the 4~\% sample in the lower panel.  Heat capacity is a particularly useful and important characterization technique as it can be done in zero field.  Previous magnetization results using a sensitive SQUID magnetometer have shown that even a small applied magnetic field can have a large effect on the spin-Peierls transition.~\cite{Hase93:70}

	Figure~\ref{hc} shows that the 2~\% sample displays the presence of two  transitions, the higher temperature transition associated with the  formation of the  spin-Peierls phase, and the low temperature transition is the formation of  AF order.  The spin-Peierls phase can be well understood in  terms of BCS theory with a gap opening in the excitation spectrum.  Therefore, the heat capacity should show a discontinuity with an exponential decay at low temperatures.   This behavior is qualitatively seen in Fig.~\ref{hc} below $T_{SP} \sim 13$~K but due to the limited temperature range below the spin-Peierls transition a  detailed fit was not possible.  Because of the presence of impurities, the spin-Peierls  transition is  broadened compared to the pure material.~\cite{Oseroff95:74}  The lower  panel of  Fig.~\ref{hc} shows heat capacity results for the 4~\% sample which shows no sign of a spin-Peierls transition at high temperatures but a clear transition,  associated with the formation of long-range AF order, is  observed at  low temperatures.  It is interesting to note from Fig.~\ref{hc} that the  2~\% sample  has a much broader N\'eel transition than in the 4~\% sample and points to a more disordered ground state in the 2~\% sample than the 4~\% sample.   This  assertion will be discussed in more detail in Sec. V.  

	The high temperature values of $C_{p}$ in both the 2~\% and 4~\% samples agree  very well and with those of previous heat capacity measurements for samples with similar dopants.~\cite{Oseroff95:74}   We have fit the high temperature heat capacity data to the form $C_{p}=\gamma T+\beta_{1}T^{3}+\beta_{2}T^{5}$, the same model  used by Oseroff \textit{et al.}  The term linear in temperature, $\gamma T$,  is  characteristic of a 1D AF systems with $\gamma=2Nk_{B}^{2}/3J$.~\cite{Wei77:21}  The term proportional to $T^{3}$ is  associated with the phonon contribution to the heat capacity.  Fitting this equation to the high temperature heat capacity data for the 4~\% sample gives  $\gamma=47 mJ/mol K^{2}$ implying $J \sim 10$~meV in good agreement with  previous measurements of the superexchange along the [001] direction and  previous  heat capacity data.  
	Based on the combination of heat capacity and magnetization we have extracted the spin-Peierls transition temperature of $T_{SP} = 11$~K for the 2~\% sample and N\'eel temperatures of $T_{N}$ = 2.5 K and 4.3 K for the 2~\% and 4~\% samples, respectively.  These values are in good agreement with the comprehensive study of Masuda $\textit{et al.}$ indicating the good quality and homogeneity of the samples.~\cite{Masuda98:80}  

\begin{figure}[t]
\begin{center}
\includegraphics[width=8.5cm] {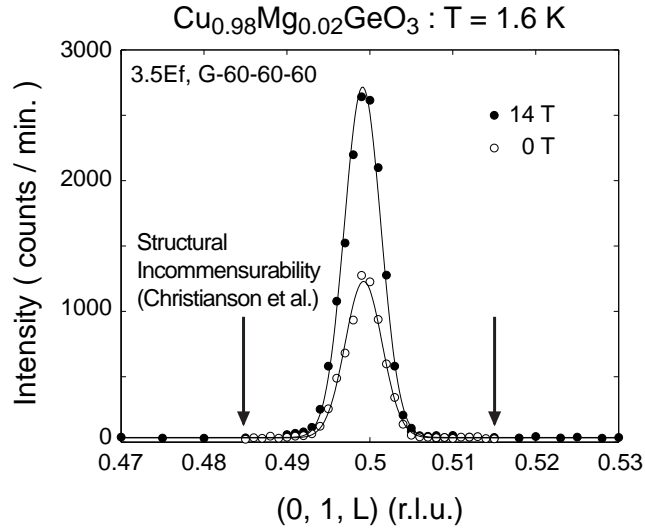}
\end{center}
\caption{\label{cold_bragg} Transverse scan through the magnetic Bragg 
peak at applied fields of 0 and 14 T along the [100] direction.  The peak 
positions of the structural incommensurate peaks are marked by arrows.}
\end{figure}

\section{Neutron Scattering Results}

\subsection{Commensurate Magnetic Peak in 2~\% Sample}

	Cold neutron diffraction was performed to study the field dependence of the magnetic order in the 2~\% sample.  The results are summarized in Fig.~\ref{cold_bragg} showing the magnetic Bragg peak profile scanned along the $c^{*}$ direction at 1.6~K with $\mu_{0}H=0$~T and 14~T, well above the commensurate to incommensurate phase transition.  One of the most striking features is the large increase in intensity at high magnetic fields.  This will be discussed in more detail in the next section.

	Another surprising result illustrated in Fig.~\ref{cold_bragg} is the fact that the magnetic Bragg peak remains \textit{commensurate} in the high field structural incommensurate phase.  A Gaussian fit to the data shows no sign of any broadening as a function of applied field.  X-ray studies of Mg-doped CuGeO$_{3}$ have shown a large splitting of the structural superlattice Bragg peaks into incommensurate positions along the $c*$ direction indicating the formation of antiphase domains characteristic of a breaking of the dimer bonds.~\cite{Kiryukhin96:76} Mg-doped CuGeO$_{3}$ with an impurity concentration of $\sim 2$~\% has been studied using high resolution synchrotron x-ray scattering and the position of the incommensurate structural Bragg peaks at 11~T are marked in Fig.~\ref{cold_bragg}.  Christianson \textit{et al.}~\cite{Christianson02:66} measured the distance between the incommensurate peaks, defined as the incommensurability parameter $\delta$, to be $\sim 0.014$~rlu  at 11~T.  Since the  spins couple antiferromagnetically, if the magnetic structure also formed antiphase domains with the same spatial modulation as the structure the magnetic Bragg peak should also split with $\delta \sim 0.007$ along the  same direction.  It is clear from Fig.~\ref{cold_bragg} that no such  splitting or broadening is present in the incommensurate phase.  This is a  very surprising result as this suggests that the spatial moment  distribution within the sample is not strongly coupled to the bulk structure.  Further discussion of this point in the context of impurities  will be presented later.  

\begin{figure}[t]
\begin{center}
\includegraphics[width=8cm] {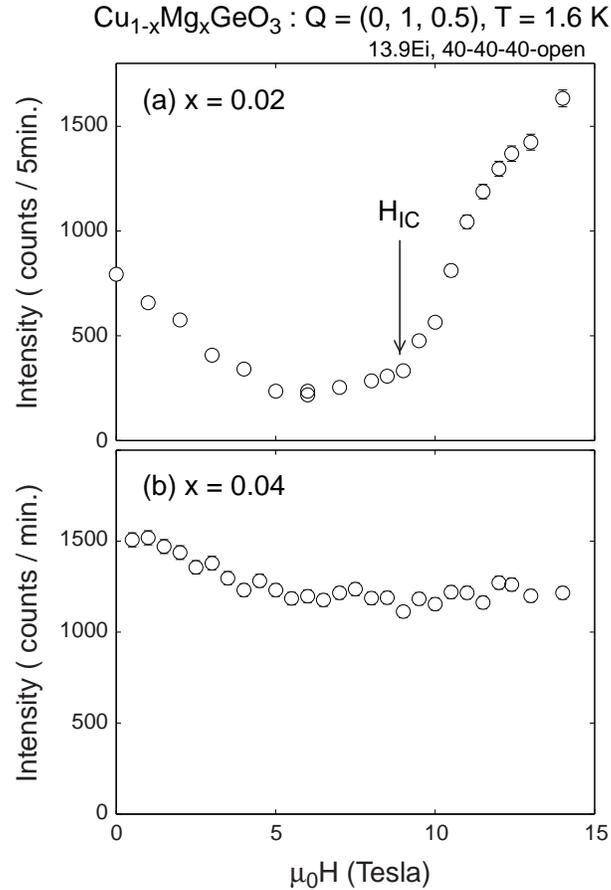}
\end{center}
\caption{\label{bragg_H} The Bragg peak intensity plotted as a function of 
applied magnetic field for both the 2~\% and 4~\% samples.  The 
enhancement of the magnetic Bragg scattering clearly occurs at the 
incommensurate transition at $\sim$ 9.5 T.}
\end{figure}

\begin{figure}[t]
\begin{center}
\includegraphics[width=8cm] {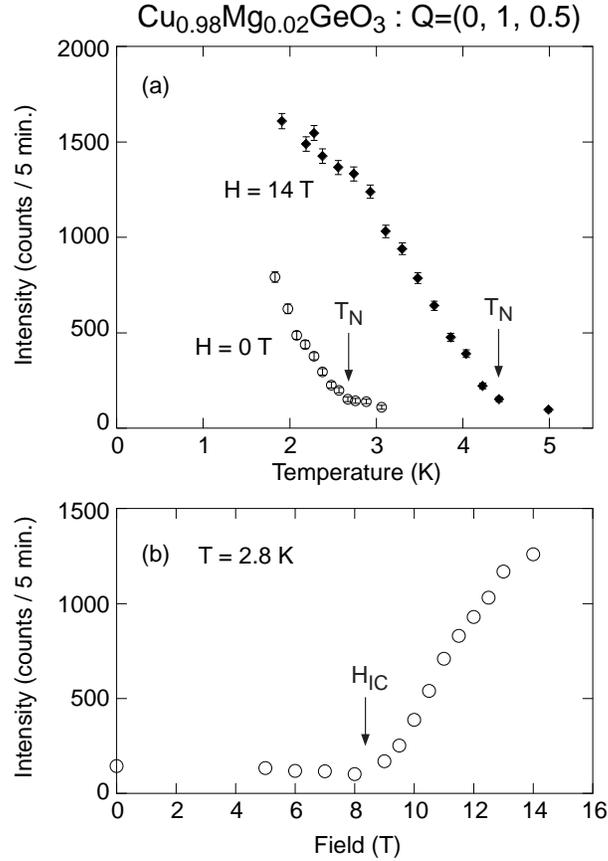}
\end{center}
\caption{\label{Fig1} The upper panel plots the magnetic Bragg peak 
intensity as a function of temperature for both 0 and 14 T. A large 
increase in the N\'eel temperature is observed.  The lower panel plots the 
magnetic Bragg peak intensity at 2.8 K as a function of field showing that the
increase in the N\'eel temperature starts at around the critical field.}
\end{figure}

\begin{figure}[t]
\begin{center}
\includegraphics[width=8cm] {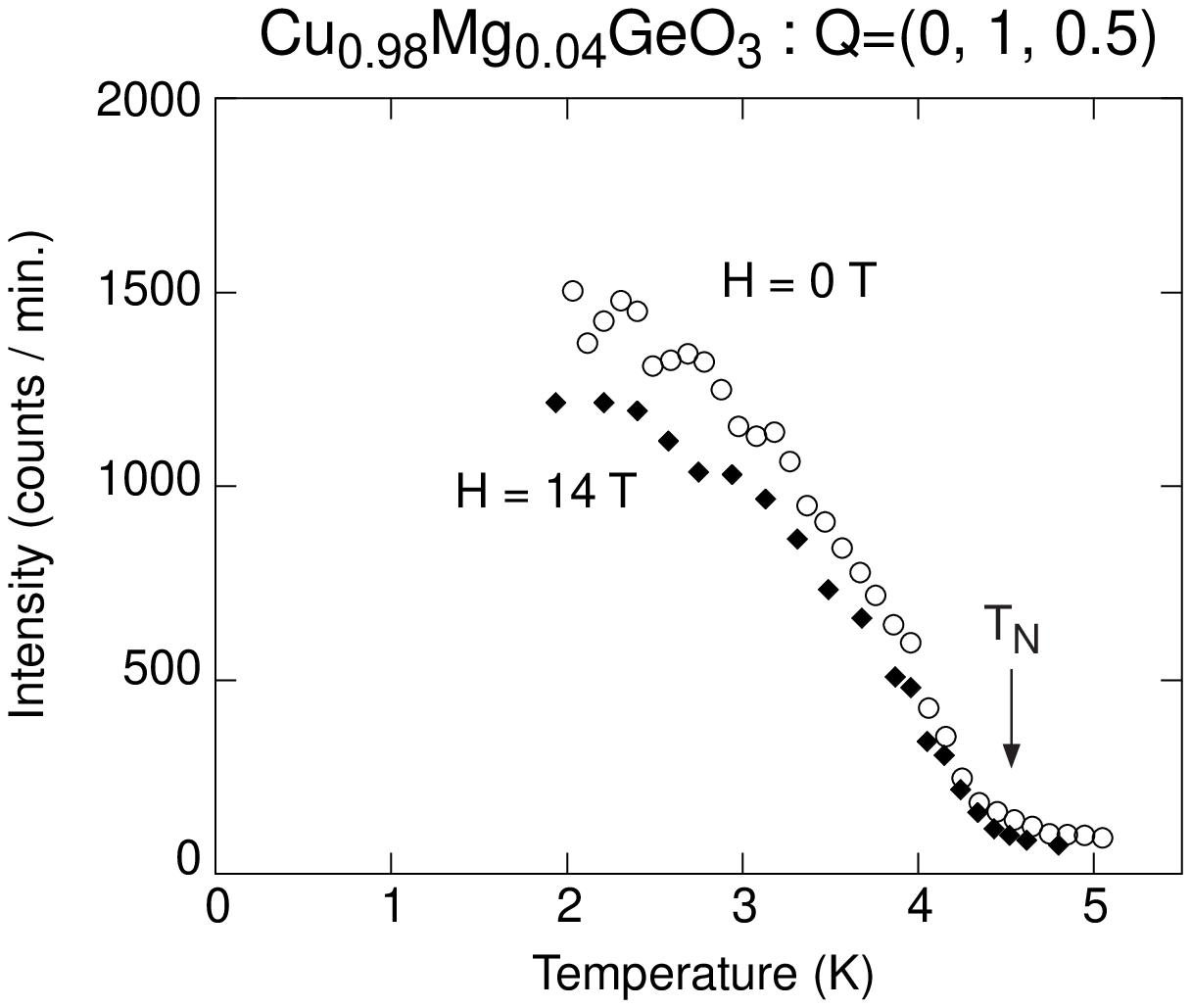}
\end{center}
\caption{\label{Fig2} The magnetic Bragg peak intensity plotted as a 
function of temperature showing the N\'eel temperature at both 0 and 14 T.  
There is no change in the N\'eel temperature for the x=0.04 sample as a function of field.}
\end{figure}

\subsection{Enhancement of Magnetic Bragg Scattering in the Incommensurate Phase}

	To investigate the enhancement of the magnetic Bragg peak illustrated in Fig.~\ref{cold_bragg} we have studied the intensity of the magnetic Bragg peaks as a function of applied magnetic field for both the 2~\% and 4~\% sample using thermal neutrons.  The essential results at low temperatures for the two dopings studied is summarized in Fig.~\ref{bragg_H}. For the dimerized AF 2~\% sample a suppression of the peak intensity is observed followed by a significant enhancement of the peak intensity at the incommensurate transition.  This is contrast to the  uniform AF 4\% sample, displayed in Fig.~\ref{bragg_H} (b) which only shows a suppression at low fields and a relatively constant behavior at higher  fields.   This comparison unambiguously shows that the enhancement is associated  with  the high-magnetic field incommensurate phase.

	The suppression of the peak intensity at intermediate fields is clearly present for both the 2~\% and 4~\% samples and therefore is not unique to the dimerized antiferromagnet.  A possible origin of this suppression was suggested by Grenier \textit{et al.}~\cite{Grenier98:44} based on magnetization data.  They proposed that the attenuation in Bragg intensity was due to the creation of free spins around the impurities.  We observe a larger fractional suppression of the AF Bragg peak intensity for the 2~\% sample than the 4~\% sample.  In the context of the model proposed by Grenier \textit{et al.} this suggests that magnetic order is largely associated with the impurities and therefore is more delicate in the dimerized AF 2~\% sample.  A similar model has recently been proposed by Glazkov \textit{et al.} to explain ESR data taken at low-magnetic fields.~\cite{Glazkov04:xx}
   
	We have determined the Cu$^{2+}$ moment direction in both the commensurate and incommensurate phases by measuring the integrated magnetic intensities at the $(0, 1, 0.5)$, $(0, 1, 1.5)$ and $(0, 3, 0.5)$ positions in each of the four quadrants.  As noted earlier, the introduction of impurities into CuGeO$_{3}$ has been found to alter the spin ground state and therefore it is important to see if the moment direction changes as a function of magnetic field.  In the commensurate phase we find the moment direction to be along [001], consistent with our low-field DC magnetization, and we find no change in the incommensurate phase.  Therefore the enhancement cannot be accounted for by a change in the moment direction. 

	The top panel of Fig.~\ref{Fig1} plots the AF peak intensity as a function of temperature for the 2~\% sample at applied magnetic fields of 0~T and 14~T.  This illustrates that the enhanced scattering is associated with a large increase in the N\'eel temperature.  We have checked this result for other incident wavelengths eliminating the possibility of it resulting from a spurious process such as multiple scattering.  The lower panel of Fig.~\ref{Fig1} plots the peak intensity at $T=2.8$~K, just above the zero field N\'eel temperature, as a function of applied magnetic field.  There is a clear break followed by an  enhancement at the incommensurate transition again showing that the enhancement of magnetic order is connected directly to the incommensurate phase transition.  These results for the 2~\% sample contrast to the field dependence of the N\'eel temperature for the 4~\% sample displayed in Fig.~\ref{Fig2} where no change in the N\'eel temperature is observed at low and high magnetic fields.  This comparison also demonstrates that the enhancement of the magnetic Bragg peak is associated with the commensurate-incommensurate transition.

	This result is consistent with the high resolution synchrotron x-ray data  of Christianson \textit{et al.} Even though x-rays are not directly sensitive to the N\'eel order it was noted in that experiment that the N\'eel temperature could be measured by a broadening of the superlattice structural Bragg peaks associated with the spin-Peierls phase at low temperature.  The high magnetic field data are suggestive of an increase in the N\'eel temperature, the result we have found here.

	It is also interesting to note that the low temperature Bragg peak intensity in the 2~\% sample does not saturate by 1.9 K at zero field and even at high magnetic fields where we can study temperatures several degrees below the N\'eel  temperature.  This differs from the behavior of the 4~\% sample when the peak intensity appears to saturate at low temperatures.  This result may suggests a considerable amount of disorder in the dimerized antiferromagnet seen in  the  2~\% sample.  This point will be discussed further in the next section.

\section{Discussion and Conclusions} 

	The most important results of the present study are the absence of incommensurate AF order and the enhancement of the commensurate AF order in the structural incommensurate phase.  In this section we discuss these features in terms of impurity effects caused by Mg doping and soliton formation by magnetic field.

	$\mu$SR measurements have been interpreted on the basis of an inhomogeneous ordered moment distribution throughout the sample for both Zn- and Si-doped CuGeO$_{3}$.~\cite{Kojima97:79} Based on this result a model was suggested where the moment size varies strongly and is peaked around the impurity sites. Such a ground state has also been suggested based on neutron scattering results.~\cite{Hase96:65}  In a detailed study of Zn doped CuGeO$_{3}$ the ordered moment was only $\sim 0.2$~$\mu_{B}$ at 1.4~K with no sign of saturation of the magnetic Bragg intensity.  Also, overdamped magnetic excitations were  observed  around the magnetic Bragg peaks, suggesting an unusual ground state of magnetic order possibly incorporating  considerable  structural disorder.  These facts suggest that the AF ordered moment detected in the present study mainly comes from regions around the impurity sites.  Since the impurities act to break the dimer bonds at low temperatures the structure around the impurities will remain commensurate  even at high magnetic fields.  Therefore, the commensurate magnetic structure in both the structurally commensurate and incommensurate phases can be understood if the static AF order is strongly pinned and localized around impurities.

	This model is also supported by our low-field data showing a suppression of the magnetic Bragg peak intensity with increasing field.  If the magnetic order is associated with local droplets with an effective spin $S=1/2$ around impurities, then long-range order is formed when the droplets become highly correlated.  In the presence of a magnetic field the antiferromagnetic correlation between each droplet will be suppressed as the spin associated with the droplet will tend to align with the magnetic field.  The properties measured here for Mg doped CuGeO$_{3}$ bear a strong similarity to that measured in doped the Haldane material Pb(Ni$_{1-x}$Mg$_{x}$V$_{2}$O$_{8}$).~\cite{Masuda02:66}  However, in that system a complete disappearance of the Neel order was observed for applied fields of only $\sim$ 3-5 T for $x$=0.02.  Therefore, in Mg doped CuGeO$_{3}$ the antiferromagnetic droplets around the impurity are more strongly correlated than that in the Haldane system.  A possible reason for this may be the strong interchain coupling in CuGeO$_{3}$.  

	This picture also provides an elegant explanation for the enhancement of the magnetic Bragg peaks in the structural incommensurate phase, induced by high field.  In the incommensurate phase, solitons are formed which break the dimer bonds.  The solitons therefore have the analogous effect of impurities with the main difference being that the solitons are mobile.  Since impurities drive magnetic order an enhancement of the magnetic Bragg peak occurs in the incommensurate phase by the soliton formation.

	Our dimerized AF sample has an impurity concentration of 2~\%, as characterized by heat capacity and DC magnetization, therefore implying an impurity or broken chain every 50 Cu$^{2+}$ ions.  At 11~T the incommensurate splitting, $\delta$, has been measured by synchrotron x-rays to be $\sim 0.014$~rlu.~\cite{Christianson02:66} By directly applying a 1D model of antiphase domains this would imply solitons every $\sim 1/(2\delta)$ or $\sim 40$ Cu atoms.  This means the 11~T field creates a number of solitons which is equivalent to a 2.5\% impurity doping.  Therefore, at high magnetic fields, the effective impurity concentration, a summation of chemical impurity and solitons, practically doubles  essentially  pushing the impurity concentration to $\sim 4$~\%.  This argument is verified  by the fact that the N\'eel temperature of the 2\% sample in the 14~T field is identical to the N\'eel temperature for 4~\%.

	The enhancement of AF order can be understood in terms of competing order parameters.   As an example, we consider the Landau theory of Kohno \textit{et al.}~\cite{Kohno99:68} for competition between superconductivity and antiferromagnetism order parameters. It was found that a staggered moment would emerge in the  vicinity of the  impurity site where the  superconducting order parameter is suppressed.  This is exactly the case in Mg doped CuGeO$_{3}$ where the impurities are the doped Mg as well as the solitons in the high magnetic field incommensurate phase.  Due to the generality of Landau theory the superconducting order parameter in the above theory can be replaced with that for the spin-Peierls order parameter.  Thus it is straightforward to see that the impurities of Mg ions and solitons suppress the spin-Peierls order and enhance the magnetic order in this Landau theory.

	Our results may also be a direct test of theories involving impurity doped $S=1/2$ chains.  The case of impurities doped into a $S=1/2$ chains has been solved using  field theory and simulated by Monte Carlo techniques by Eggert 
and  Affleck.~\cite{Eggert92:46,Eggert95:75}  In this work it was found that the stable fixed point was an  open chain and staggered moments would appear near the open end.  This theory directly  applies to the impurity doped CuGeO$_{3}$ where the dopants or the solitons can be the open end of the chain.  For the particular case of the CuGeO$_{3}$ a solution to the spin problem  has been  suggested by Saito and Fukuyama~\cite{Saito97:66}. They found that the  staggered  magnetization could be written in terms a quantum phase variable $\theta (x)$, which is a solution of the Sine-Gordan equation,  and that magnetic  order 
would form around a {\it domain wall} of dimerization caused by an impurity. This theory can be directly applied to the presence of solitons and clearly illustrates the analogous effects of impurity doping and the formation of solitons.

	The two main results of this high field experiment are the presence of a  commensurate magnetic Bragg peak in both the low field commensurate and  the  high field incommensurate phases and the large enhancement of the magnetic  Bragg  peak in the incommensurate phase only.  We have obtained these results  from a  comparison of a dimerized sample to that of a uniform antiferromagnet by  changing the Mg concentration.  The results are consistent with solitons  formed at high magnetic fields effectively acting as impurities.  These  results  are consistent with theories based on competing order parameters and  impurity  effects in $S=1/2$ chains. 

\section*{Acknowledgment}

We would like to thank H. Graf, Y. J. Wang, and G. Shirane for useful discussions.  Work at the University of Toronto was supported by the Natural Sciences and Engineering Research Council of Canada and by the National Research Council of Canada.

\end{document}